\documentclass[a4paper,9pt, onecolumn]{article}
\usepackage{lmodern}
\usepackage[utf8]{inputenc}
\usepackage[T1]{fontenc}
\usepackage[english]{babel}
\usepackage[usenames,svgnames]{xcolor}
\usepackage{setspace}
\usepackage{endnotes}
\usepackage[labelfont=bf]{caption}
\usepackage{subcaption}
\usepackage{authblk}
\usepackage[pdftex]{graphicx} 
\usepackage[pdftex,pageanchor=true,hyperindex=true]{hyperref}
\usepackage{epstopdf} 
\usepackage{bookmark}
\usepackage{pdfpages}
\usepackage{array}
\usepackage{type1cm,soul}
\usepackage{tabularx}
\usepackage{ifthen} 
\usepackage{lscape}
\usepackage{amsmath}
\usepackage{amsfonts}
\usepackage{wasysym}
\usepackage{fullpage}
\usepackage{xstring}
\usepackage{titlesec}
\usepackage{colortbl}
\usepackage{tcolorbox}
\usepackage{calc}
\usepackage{enumitem}
\usepackage{gensymb} 
\usepackage{pifont} 
\usepackage{booktabs} 
\usepackage{multirow} 
\usepackage{bm} 
\usepackage{siunitx}

\graphicspath{{Figures/}}

\font\titlefont=cmr12 at 15pt
\title{\titlefont Sub-second photon dose prediction via transformer neural networks}

\author[1,2]{Oscar Pastor-Serrano}
\author[2]{Peng Dong}
\author[3]{Charles Huang}
\author[2*]{Lei Xing}
\author[1*]{Zoltán Perkó}

\affil[1]{\small Delft University of Technology, Department of Radiation Science \& Technology, Delft, Netherlands}

\affil[2]{\small Stanford University, Department of Radiation Oncology, Stanford, CA, USA}

\affil[3]{\small Stanford University, Department of Bioengineering, Stanford, CA, USA}

\affil[*]{\small These authors share the senior authorship}

\date{}

\begin{document}
\newcommand{\FixRef}[3][sec:]
{\IfBeginWith{#2}{#3}
	{\StrBehind{#2}{#3}[\RefResult]}
	{\def\RefResult{#2}}\IfBeginWith{#1}{#3}
	{\StrBehind{#1}{#3}[\RefResultb]}
	{\def\RefResultb{#1}}}

\newcommand{\secref}[1]
{\FixRef{#1}{sec:}Section~\ref{sec:\RefResult}}
\newcommand{\secreff}[1]
{\FixRef{#1}{sec:}in Section~\ref{sec:\RefResult}}
\newcommand{\Secreff}[1]
{\FixRef{#1}{sec:}In Section~\ref{sec:\RefResult}}
\newcommand{\secrefm}[2]
{\FixRef[#2]{#1}{sec:}Sections~\ref{sec:\RefResult}-\ref{sec:\RefResultb}}
\newcommand{\secreffm}[2]
{\FixRef[#2]{#1}{sec:}in Sections~\ref{sec:\RefResult}-\ref{sec:\RefResultb}}
\newcommand{\Secreffm}[2]
{\FixRef[#2]{#1}{sec:}In Sections~\ref{sec:\RefResult}-\ref{sec:\RefResultb}}
\newcommand{\figref}[1]
{\FixRef{#1}{fig:}Figure~\ref{fig:\RefResult}}
\newcommand{\figrefm}[2]
{\FixRef[#2]{#1}{fig:}Figures~\ref{fig:\RefResult}-\ref{fig:\RefResultb}}
\newcommand{\figreff}[1]
{\FixRef{#1}{fig:}in Figure~\ref{fig:\RefResult}}
\newcommand{\figreffm}[2
]{\FixRef[#2]{#1}{fig:}in Figures~\ref{fig:\RefResult}-\ref{fig:\RefResultb}}
\newcommand{\Figreff}[1]
{\FixRef{#1}{fig:}In Figure~\ref{fig:\RefResult}}
\newcommand{\Figreffm}[2]
{\FixRef[#2]{#1}{fig:}In Figures~\ref{fig:\RefResult}-\ref{fig:\RefResultb}}
\newcommand{\tabref}[1]
{\FixRef{#1}{tab:}Table~\ref{tab:\RefResult}}
\newcommand{\tabreff}[1]
{\FixRef{#1}{tab:}in Table~\ref{tab:\RefResult}}
\newcommand{\Tabreff}[1]
{\FixRef{#1}{tab:}In Table~\ref{tab:\RefResult}}
\newcommand{\tabrefm}[2]
{\FixRef[#2]{#1}{tab:}Tables~\ref{tab:\RefResult}-\ref{tab:\RefResultb}}
\newcommand{\tabreffm}[2]
{\FixRef[#2]{#1}{tab:}in Tables~\ref{tab:\RefResult}-\ref{tab:\RefResultb}}
\newcommand{\Tabreffm}[2]
{\FixRef[#2]{#1}{tab:}In Tables~\ref{tab:\RefResult}-\ref{tab:\RefResultb}}
\newcommand{\egyref}[1]
{\FixRef{#1}{eq:}Equation~\ref{eq:\RefResult}}
\newcommand{\eqreff}[1]
{\FixRef{#1}{eq:}in Equation~\ref{eq:\RefResult}}
\newcommand{\Eqreff}[1]
{\FixRef{#1}{eq:}In Equation~\ref{eq:\RefResult}}
\newcommand{\eqrefm}[2]
{Equations~\ref{eq:#1}-\ref{eq:#2}}
\newcommand{\eqreffm}[2]
{\FixRef[#2]{#1}{eq:}in Equations~\ref{eq:\RefResult}-\ref{eq:\RefResultb}}
\newcommand{\Eqreffm}[2]
{\FixRef[#2]{#1}{eq:}In Equations~\ref{eq:\RefResult}-\ref{eq:\RefResultb}}
\newcommand{\charef}[1]
{\FixRef{#1}{cha:}Chapter~\ref{cha:\RefResult}}
\newcommand{\chareff}[1]
{\FixRef{#1}{cha:}in Chapter~\ref{cha:\RefResult}}
\newcommand{\Chareff}[1]
{\FixRef{#1}{cha:}In Chapter~\ref{cha:\RefResult}}

\newcommand*{\dd}{\mathrm{d}}
\newcommand{\ui}[1]{\textit{\textbf{#1}}}
\newcommand{\mx}[1]{\underline{\underline{#1}}}
\newcommand{\diff}[2]{\dfrac{\dd #1}{\dd #2}}
\newcommand{\pdiff}[2]{\dfrac{\partial #1}{\partial #2}}
\newcommand{\dhl}{\hline\hline}
\newcommand{\rb}[1]{\left(#1\right)}
\newcommand{\sqb}[1]{\left[#1\right]}
\newcommand{\tb}[1]{\left<#1\right>}
\newcommand{\cb}[1]{\left\{#1\right\}}
\newcommand{\abs}[1]{\left|#1\right|}
\newcommand{\dspm}[1]{\begin{displaymath}#1\end{displaymath}}
\newcommand{\ds}{\displaystyle}
\newcommand{\pow}[2]{\cdot #1^{#2}}
\newcommand{\evat}[2]{\left.#1\right|_{#2}}
\newcommand{\ifrac}[2]{\ds #1 / #2}
\newcommand{\ab}{\ifrac{\alpha}{\beta}}
\newcommand{\BED}{\text{BED}}
\newcommand{\norm}[1]{\left\lVert#1\right\rVert}

\newcommand{\cmark}{\ding{51}}%
\newcommand{\xmark}{\ding{55}}%

\def\thetable{\Roman{table}}
\thispagestyle{empty}
\onecolumn
\maketitle
\noindent

\begin{abstract}
\noindent\textbf{Background:} Fast dose calculation is critical for online and real time adaptive therapy workflows. While modern physics-based dose algorithms must compromise accuracy to achieve low computation times, deep learning models can potentially perform dose prediction tasks with both high fidelity and speed. 

\noindent\textbf{Purpose:} We present a deep learning algorithm that, exploiting synergies between Transformer and convolutional layers, accurately predicts broad photon beam dose distributions in few milliseconds. 
 
\noindent\textbf{Methods:} The proposed improved Dose Transformer Algorithm (iDoTA) maps arbitrary patient geometries and beam information (in the form of a 3D projected shape resulting from a simple ray tracing calculation) to their corresponding 3D dose distribution. Treating the 3D CT input and dose output volumes as a sequence of 2D slices along the direction of the photon beam, iDoTA solves the dose prediction task as sequence modeling. The proposed model combines a Transformer backbone routing long-range information between all elements in the sequence, with a series of 3D convolutions extracting local features of the data. We train iDoTA on a dataset of 1700 beam dose distributions, using 11 clinical volumetric modulated arc therapy (VMAT) plans (from prostate, lung and head and neck cancer patients with 194-354 beams per plan) to assess its accuracy and speed.  

\noindent\textbf{Results:} iDoTA predicts individual photon beams in $\approx50$ milliseconds with a high gamma pass rate of $97.72\pm1.93\%$ (2 mm, 2\%). Furthermore, estimating full VMAT dose distributions in 6-12 seconds, iDoTA achieves state-of-the-art performance with a $99.51\pm0.66\%$ (2 mm, 2\%) pass rate and an average relative dose error of $0.75\pm0.36$\%. 

\noindent\textbf{Conclusions:} Offering the sub-second speed needed in online and real-time adaptive treatments, iDoTA represents a new state of the art in data-driven photon dose calculation. The proposed model can massively speed-up current photon workflows, reducing calculation times from few minutes to just a few seconds.
\end{abstract}

\section{Introduction}
\label{sec:Introduction}
Modern radiotherapy techniques such as intensity modulated radiation therapy (IMRT) or volumetric modulated arc therapy (VMAT) critically rely on accurate and fast calculations of the radiation dose delivered within the patient by photon beams, typically shaped by multi-leaf collimators (MLC)  \cite{hussein_automation_2018}. With modern workflows moving towards online or real time adaptation, fast dose calculations are critical for quick plan evaluation, re-optimization and finally being able to account for motion due to breathing or anatomical changes.

Commercial treatment planning systems mainly use pencil beam (PB) \cite{mohan_differential_1986}, collapsed cone (CC) \cite{boyer_photon_1985, ahnesjo_collapsed_1989}, or Monte Carlo (MC) dose engines. While both PB and CC algorithms are usually faster than MC, the assumptions and approximations they use to solve photon particle transport result in less accurate results. Conversely, MC methods --- the gold standard in dose calculation --- simulate individual stochastic particle trajectories abiding the physical laws of nuclear interactions and track the deposited dose along these paths. By averaging results from enough particles (typically several millions), MC methods achieve very high accuracy even in the most complex patient geometries, at the cost of high computation times. Current commercial treatment planning systems mainly use improved PB or CC variations yielding close-to-MC accuracy, e.g., the anisotropic analytical algorithm (AAA) \cite{ulmer_3d_2005, sievinen_aaa_nodate} based on the PB convolution \cite{mohan_differential_1986} in Eclipse (Varian Medical Systems) or the CC convolution algorithm in Pinnacle (Philips) \cite{boyer_photon_1985}. Some recent MC implementations also use the parallelization capabilities of graphics processing units (GPUs) to reduce dose calculation times from several hours to minutes \cite{jia_gpu-based_2011,jahnke_gmc_2012, hissoiny_fast_2011}. Despite these advances, the need for accurate and fast dose calculation algorithms is still unmet in most clinical workflows, as neither PB nor MC are fast enough for real time treatment plan correction. 

Recently, deep learning models have been applied to several steps of the radiotherapy workflow \cite{meyer_survey_2018}, mainly as U-net convolutional architectures \cite{ronneberger_u-net_2015} or Generative Adversarial Networks \cite{goodfellow_generative_2014}. Most works aim to aid treatment planning by predicting clinically optimal doses based on historical data. As a result, they are constrained to a specific site, clinical optimum choice, and often fixed beam configurations, limiting their generalization capabilities. These models typically directly predict the full dose distribution using computed tomography (CT) images \cite{kearney_dosenet_2018}, organ masks \cite{chen_feasibility_2019, fan_automatic_2019, nguyen_feasibility_2019, kajikawa_convolutional_2019, ma_dose_2019}, or additional information about the photon beam configuration \cite{nguyen_3d_2019} as input. To further aid treatment planning, few studies additionally provide the beam intensities needed to deliver the predicted dose distribution \cite{lee_fluence-map_2019, wang_fluence_2020}. 

Aiming at predicting dose distributions in generic setups, several subsequent studies present dose calculation models that estimate beam or full dose distributions from CTs and additional physics input such as high noise MC \cite{ peng_mcdnet_2019, bai_deep_2021, neph_deepmc_2021} or PB doses \cite{xing_boosting_2020, dong_deep_2020}; fluence maps, e.g., resulting from simple ray tracing calculations \cite{fan_data-driven_2020, xing_technical_2020}; energy released per unit mass \cite{zhu_preliminary_2020}; or a combination of the previous with additional beam information \cite{kontaxis_deepdose_2020, tsekas_deepdose_2021}. The reason for their success are the convolutional layers that excel at capturing local features and are heavily optimized for GPU hardware, but are less appropriate for modeling long-range dependencies, e.g., changes along the beam direction through the patient.

Although some of the most recent models can quickly predict dose distributions in most cases with good accuracy \cite{kontaxis_deepdose_2020, tsekas_deepdose_2021}, there is room for improvement with newer architectures that require less input information and can model distant features in the data. Recent Transformer architectures \cite{vaswani_attention_2017} are particularly well-suited to process local and distant features, yielding excellent results in a wide range of sequence modeling tasks \cite{devlin_bert_2019,brown_language_2020,dosovitskiy_image_2020}. For smaller datasets, Transformers perform particularly well when combined with convolutional layers \cite{dascoli_convit_2021}. Based on these synergies between convolutions and Transformers, a recent study presented a transformer-based algorithm predicting proton beamlet 3D dose distributions as a sequence of 2D slices in the beam depth, with state-of-the-art performance and speed \cite{pastor-serrano_learning_2022, pastor-serrano_millisecond_2022}. 

\begin{figure*}[t!]
	\centering
	\includegraphics[width=0.99\textwidth]{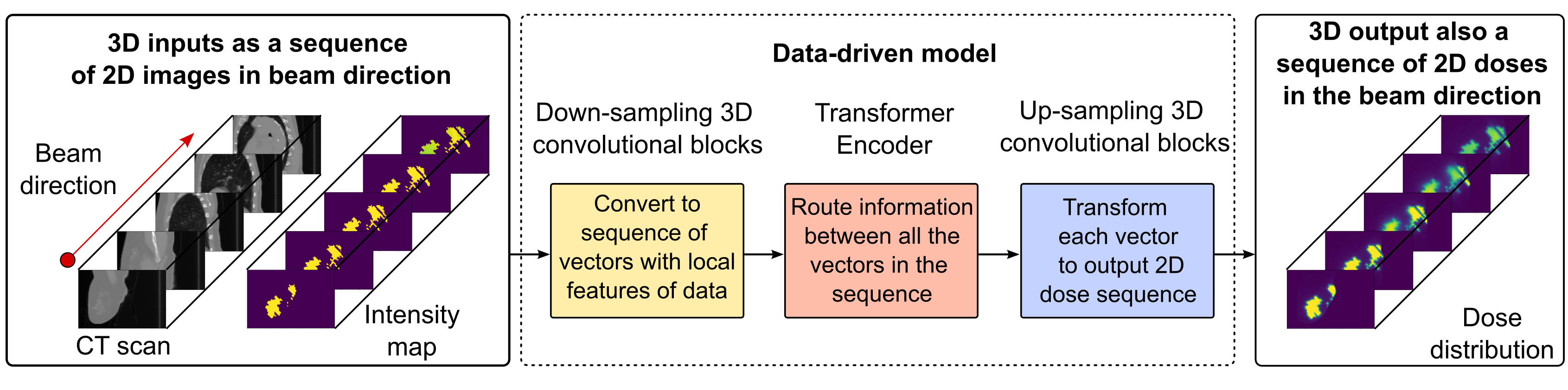} 
	\caption{\textbf{Model overview.} A deep learning data-driven model learns the mapping $\bm{y} = f_{\bm{\theta}}(\bm{x}, \bm{r})$ between input 3D CT $\bm{x}$ and projected shape $\bm{r}$ volumes, and the corresponding output 3D dose distributions $\bm{y}$. The problem is formulated as a sequence prediction task, where all input and output cubes are treated as a sequence of 2D slices in the beam's eye view. Each 2D slice is mapped into a vector via a series of down-sampling convolutional blocks. A transformer backbone routes information between all elements of the resulting sequence. Finally, a several up-sampling convolutional operations transform each vector into a 2D dose distribution map.}
	\label{fig:Model}
\end{figure*}

In this study, we present a deep learning model that can predict dose distributions in few milliseconds with clinically acceptable accuracy. As in concurrent work \cite{xiao_transdose_2022}, we harness the power of hybrid Transformer and 3D convolutional architectures, adapting the previous transformer-based proton dose calculation model \cite{pastor-serrano_learning_2022} to predict the dose of much bigger photon broad beams. As shown in \figref{Model}, the proposed improved Dose Transformer Algorithm (iDoTA) combines a series of 3D convolutional layers modeling local dose and tissue variations, with a Transformer backbone routing information along the depth of the entire photon beam. The model treats input 3D CT and projected shape volumes (containing beam geometrical information) as a sequence of 2D slices in the direction of the beam, framing dose calculation as sequence modeling to produce a sequence of 2D dose slices forming the 3D dose distribution. After comparing iDoTA to the best-performing data-driven models, we demonstrate its superior speed and accuracy for photon dose calculation tasks, being capable to speed up beam prediction times down to few milliseconds and reducing treatment plan computation times to few seconds. 

\section{Methods and materials}
\label{sec:Methods}
In this section, we present the problem setup and architecture of the iDoTA model, used to predict photon beam doses from 3D CT and projected shape inputs. Additionally, we describe the dataset and training procedure used to optimize the model parameters, as well as the evaluation metrics used to assess iDoTA's performance as a generic photon dose calculation engine.
 
\paragraph{Proposed framework} Photon dose calculation involves estimating the radiation dose delivered in the patient geometry. If the machine parameters do not change, the predicted dose distribution mainly depends on the irradiated geometry and the beam geometrical information such as the MLC aperture shape, the beam angle and the relative position of the isocenter. We assume that all the necessary beam shape information is captured in a 3D projected shape $\bm{r}\in\mathbb{R}^{D\times H\times W}$ of depth $D$, height $H$ and width $W$, containing the result of a simple ray tracing operation propagating the photon beam shape through the patient geometry CT scan $\bm{x}\in\mathbb{R}^{D\times H\times W}$. The outcome of the dose calculation operation predicted by our model is another grid $\bm{y}\in\mathbb{R}^{D\times H\times W}$ with the 3D distribution of dose per monitor unit (MU).

\begin{figure*}[t]
	\centering
	\includegraphics[width=0.99\textwidth]{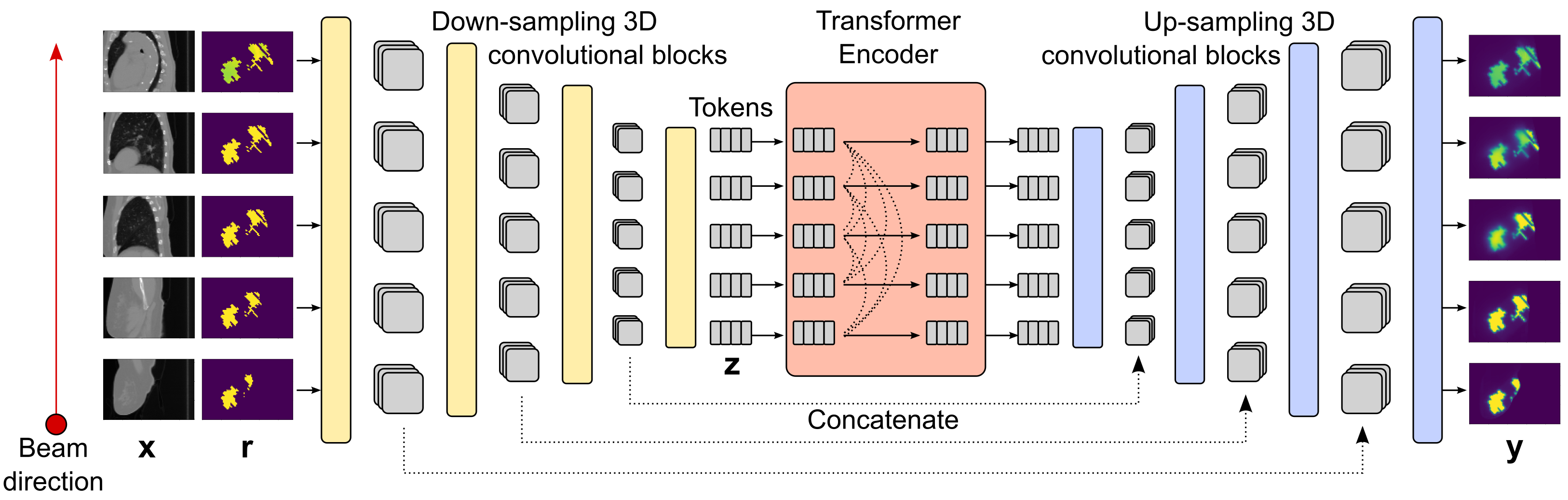} 
	\caption{\textbf{Model architecture}. The proposed model solves the dose prediction task as sequence modeling, mapping two input sequences of 2D CT slices $\bm{x}$ and projected shapes $\bm{r}$ with beam shape information  into a sequence of 2D dose distributions $\bm{y}$. First, a series of down-sampling convolutional blocks merges and compresses the two sequences from the data into a sequence of feature vectors $\bm{z}$ (referred to as tokens). A Transformer encoder with causal self-attention routes long-range dependencies along the beam direction. Finally, a series of up-sampling convolutional blocks transform the output tokens into a sequence of 2D dose distributions. In each block, the exact same 3D convolution operation is applied to all sequence elements, extracting local features from the  preceding and following element in the sequence.}
	\label{fig:Architecture}
\end{figure*}

As shown in \figref{Architecture}, the patient CT $\bm{x}$ and the 3D projected shape $\bm{r}$ are inputs to iDoTA, which during training implicitly learns the mapping $\bm{y} = f_{\bm{\theta}}(\bm{x},\bm{r})$ via a cascade of neural networks layers with parameters $\bm{\theta}$. Framing the dose prediction task as modeling a sequence of $D$ elements in the direction of the photon beam, we combine the strengths of both convolutional and Transformer architectures into a single model. The input geometry $\bm{x}$ can be expressed a sequence of $D$ images in the direction of the beam $\{\bm{x}_i|\bm{x}_i\in\mathbb{R}^{1\times H\times W},\forall i=1,...,D \}$, while the projected shape 3D input $\bm{r}$ is similarly viewed as a sequence 2D slices $\{\bm{r}_i|\bm{r}_i\in\mathbb{R}^{1\times H\times W},\forall i=1,...,D \}$ containing beam information. Likewise, the final dose volume $\bm{y}$ is also expressed as the sequence of 2D dose slices  $\{\bm{y}_i|\bm{y}_i\in\mathbb{R}^{1\times H\times W},\forall i=1,...,D \}$). 

\paragraph{Model architecture} As seen in \figref{Architecture}, the proposed architecture combines a series of convolutional blocks modeling local features with a Transformer backbone that processes information along the entire beam depth. 

\begin{itemize}
	\item First, a series of down-sampling convolutional blocks extract local features of the data into a sequence of vectors $\{\bm{z}_i|\bm{z}_i\in\mathbb{R}^{N},\forall i=1,...,D \}$ --- referred to as tokens in the remainder of the paper --- of size $N$. Each block contains a 3D convolutional layer with kernel size equal to 3, modeling local features from the immediately preceding and succeeding elements in the sequence, followed by a layer normalization \cite{ba_layer_2016}, a rectified linear unit (ReLU) activation function and a max-pooling operation. All such operations in the block are applied in parallel to every element of the input sequence. Due to the max-pooling operation, the height $H$ and width $W$ of the slices are halved after each block. A total of $L$ blocks result in $L$ resolution levels. After the last block, we apply a final convolution with $K$ filters and flatten the resulting features into tokens of dimension $N=\Big(\frac{H}{2}\Big)^L\times\Big(\frac{W}{2}\Big)^L\times K$. As a result, we obtain a sequence of $D$ tokens containing local features about the corresponding input slices, e.g., the third token $\bm{z}_3$ represents local features from the inputs $\bm{x}_3$, $\bm{r}_3$ and their neighboring slices.
	
	\item A Transformer backbone routes information between the extracted features along the depth of the entire volume, with the self-attention mechanism \cite{vaswani_attention_2017} making the information exchange dynamic, i.e., each token $\bm{z}_i$ is independently transformed based on its content and information selectively gathered from other sequence elements. To account for the relative distance between tokens, we add a learnable positional embedding $\bm{p}_i\in\mathbb{R}^N$ to each token $\bm{z}_i$, i.e., a sequence of vectors $\{\bm{p}_i|\bm{p}_i\in\mathbb{R}^{N},\forall i=1,...,D \}$ is learned and added to the token sequence before the first operation in the Transformer. We use the pre-Layer Normalization architecture \cite{xiong_layer_2020}, which consists of a Layer Normalization (LN) \cite{ba_layer_2016} operation, followed by a self-attention operation \cite{vaswani_attention_2017}, and two fully-connected layers with Dropout \cite{srivastava_dropout_2014} and a Gaussian Error Linear Unit (GeLU) activation \cite{hendrycks_gaussian_2016}. 
	
	\item Finally, a series of $L$ up-sampling convolutional blocks convert the token sequence into the output dose volume. For each level, we append (along the feature dimension) the sequence previously obtained from the same level down-sampling convolutional block, similar to U-net type architectures. The up-sampling block's architecture is identical to that of its down-sampling counterpart, except for the use of a nearest-neighbor up-sampling interpolation operation instead of the max-pooling.
\end{itemize}

\paragraph{Projected shape and dose calculation} Apart from the values in the CT, the additional 3D projected shape input $\bm{r}$ encodes beam information such as the MLC aperture shape, the angle or relative distance between the isocenter and the source, including basic material information with a simple correction based on tissue densities. Such projected shape is generated via an algorithm that estimates the dose at each voxel through the percentage depth dose (PDD), corrected by an off-axis factor. The PDD is measured at 100 cm source-to-surface distance (SSD) with a $\SI{10}{\cm}\times\SI{10}{\cm}$ field size, adjusting for different SSDs using the Mayneord factor. The depth for determining the percentage dose is the water equivalent distance, calculated via ray tracing for all voxels. The off-axis correction factor is calculated by sampling from a diagonal beam profile for a $\SI{40}{\cm}\times\SI{40}{\cm}$ field size at 10 cm depth, projecting it to different depths using the lateral distance of the voxel to the center beam axis and the longitudinal distance from the voxel to the source. This ray tracing calculation estimates the dose using the commissioning data and is optimized for speed over accuracy, taking around 0.1 ms per beam in a GPU. The corresponding ground truth dose distributions (to be predicted by the model) are obtained via the AcurosXB V15.6.05 algorithm in the Varian Eclipse TPS system (with the option of calculating dose to medium). Both the dose and the projected shapes have similar ranges from 0 to $\approx3$, with units cGy/MU.

\paragraph{Dataset} iDoTA is trained to predict individual photon beams using a training dataset of 17 clinical patient CTs with disease sites of brain, head neck, lung, abdomen and pelvis. All CTs were recorded using a General Electrics LightSpeed CT scanner with $\SI{2.5}{\mm}\times\SI{2.5}{\mm}\times\SI{2.5}{\mm}$ resolution. For each patient, 100 different co-planar photon beams were computed, using, for each beam, a random gantry angle and an isocenter location randomly selected within the patient, and an aperture shape that was generated by randomly sampling leaf positions, keeping the couch angle fixed. After calculating the dose per MU and projecting the aperture shape, we cropped 3D CT $\bm{x}\in\mathbb{R}^{96\times 96\times 64}$, projected shape $\bm{r}\in\mathbb{R}^{96\times 96\times 64}$ and dose $\bm{y}\in\mathbb{R}^{96\times 96\times 64}$ blocks covering a volume of approximately $240\times 240\times 160 $ $\text{mm}^3$, so that the beam always travels in the same direction along the first dimension $D=96$ with angles between -45$\degree$ and 45$\degree$. All 1700 input CT volumes are normalized to the range [0,1] dividing by using the maximum value of 3,071 observed across the entire dataset. Likewise, we normalize both projected shapes and dose distributions using the maximum dose value of 3.075 cGy/MU in the dataset. During training, 10\% of the samples are set aside for validation purposes, i.e., finding the best model configuration.

We evaluate the best model using an independent test dataset of 584 beam dose distributions corresponding to a prostate and a lung patient unseen during training. Additionally, to assess iDoTA's performance in predicting full dose distributions composed of many photon beams, we obtain 11 additional clinical VMAT treatment plans with 2 arcs and 99-178 control points per arc, corresponding to 1 brain, 3 HN, 3 lung, and 4 prostate cancer patients.

\paragraph{Training details} We train iDoTA using the mean squared error as a loss function, with mini-batches of 4 samples and the layer-adaptive LAMB optimizer \cite{you_large_2019}, finding the combination of a low batch size and the LAMB optimizer to be critical for convergence. During training, we augment the dataset size via rotations (in steps of 90 $\degree$, perpendicular to the direction of the beam) and random shifts along the beam direction (shifting the entire volume up to 15 positions along the first dimension). Training consists of 10 cycles with 120 epochs/cycle, where the learning rate is set to $10^{-3}$ at the beginning of each cycle, and halved every 15 epochs.

After hyper-parameter tuning using the validation data, the best-performing model has $H=4$ transformer heads, $L=4$ levels with $K=10$ filters in the last encoder convolution. The four down-sampling operations in the encoder transform the input slices with dimensions $H=96$ and $W=64$ into tokens of size $N=H/2^4\times W/2^4\times K=240$. All training and experiments are run in a Nvidia A40{\textregistered} GPU using Tensorflow \cite{abadi_tensorflow_nodate}.

\paragraph{Evaluation metrics} For evaluation purposes, we compare iDoTA's predictions to the corresponding ground truth dose distributions in the independent test set of patients unseen during training. The main method to assess dosimetric differences is the gamma analysis \cite{low_technique_1998}, based on the intuition that two neighboring voxels with a similar dose result in equivalent biological effects. Intuitively, a voxel in the predicted dose distribution passes the gamma evaluation $\Gamma(\delta\text{ mm},\Delta\%)$ if another voxel with a similar value --- deviating less than $\Delta\%$ of the maximum dose --- is found within a sphere of radius $\delta$ mm in the ground truth dose grid. We compute three gamma evaluations $\Gamma(1\text{ mm},1\%)$,  $\Gamma(2\text{ mm},2\%)$ and  $\Gamma(3\text{ mm},3\%)$, and calculate the gamma passing rate by dividing the number of passed voxels by the total amount of eligible voxels, i.e., voxels with values within 10\% and 100\% of the maximum dose.

As an additional metric to measure explicit voxel dose differences, we compute the average absolute error $\rho$, expressed as a percentage of the maximum dose in the grid. For model predictions $\bm{y}$, and corresponding ground truth 3D dose distributions $\bm{\hat{y}}$ (both with $n_v = D\times H\times W$ voxels), the average absolute error is calculated using the $L_1$-norm as 

\begin{equation}
	 \rho = \frac{1}{n_v} \frac{\norm{\bm{y}-\bm{\hat{y}}}_{L_1}}{\max{\bm{\hat{y}}}}\times 100.
\end{equation}

\section{Results}
\label{sec:Results}
To assess iDoTA's suitability as a generic photon dose calculation tool and determine its improvements with respect to other data-driven algorithms, we compute the different evaluation metrics on the independent test data. In particular, we compare iDoTA's accuracy and speed to previous approaches when predicting both individual photon beam prediction and full dose distributions from clinical VMAT plans.

\begin{figure}[t]
	\centering
	\includegraphics[width=\textwidth]{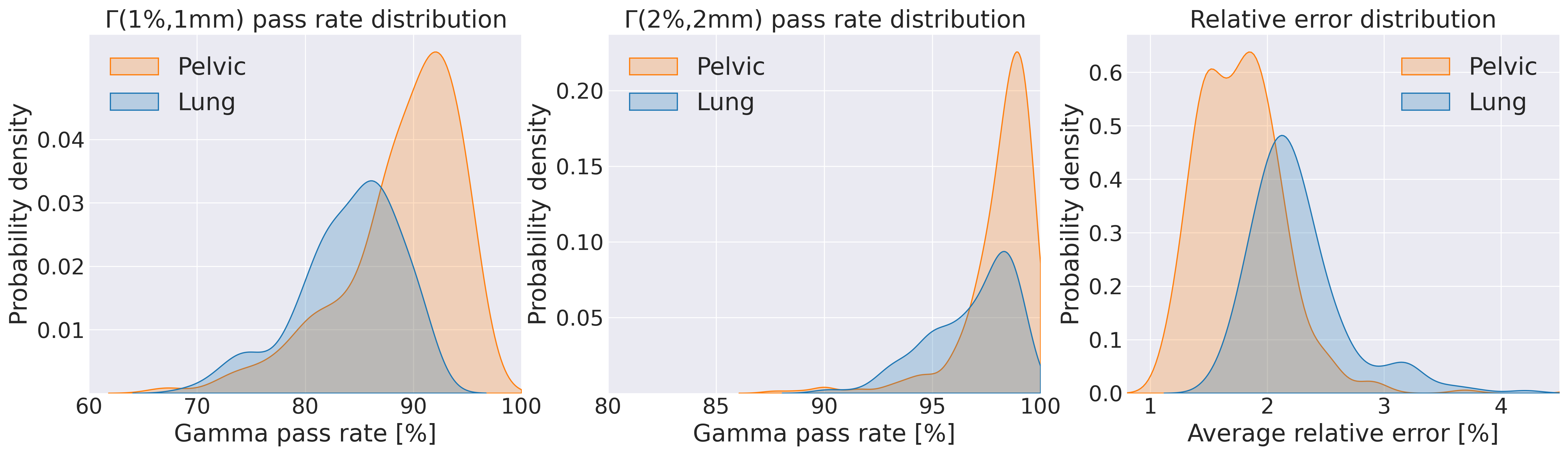}
	\caption{\textbf{Accuracy metrics distribution.} (Left)$\Gamma(1\text{ mm},1\%)$. pass rate, (middle) $\Gamma(2\text{ mm},2\%)$ pass rate and (right) absolute relative error distributions across all beams in the test dataset. The lower errors and higher pass rate values in orange correspond to beams in the pelvic area, while blue distributions are from lung samples.}
	\label{fig:beam_dist}
\end{figure}

\begin{figure}[]
	\centering
	\includegraphics[width=\textwidth]{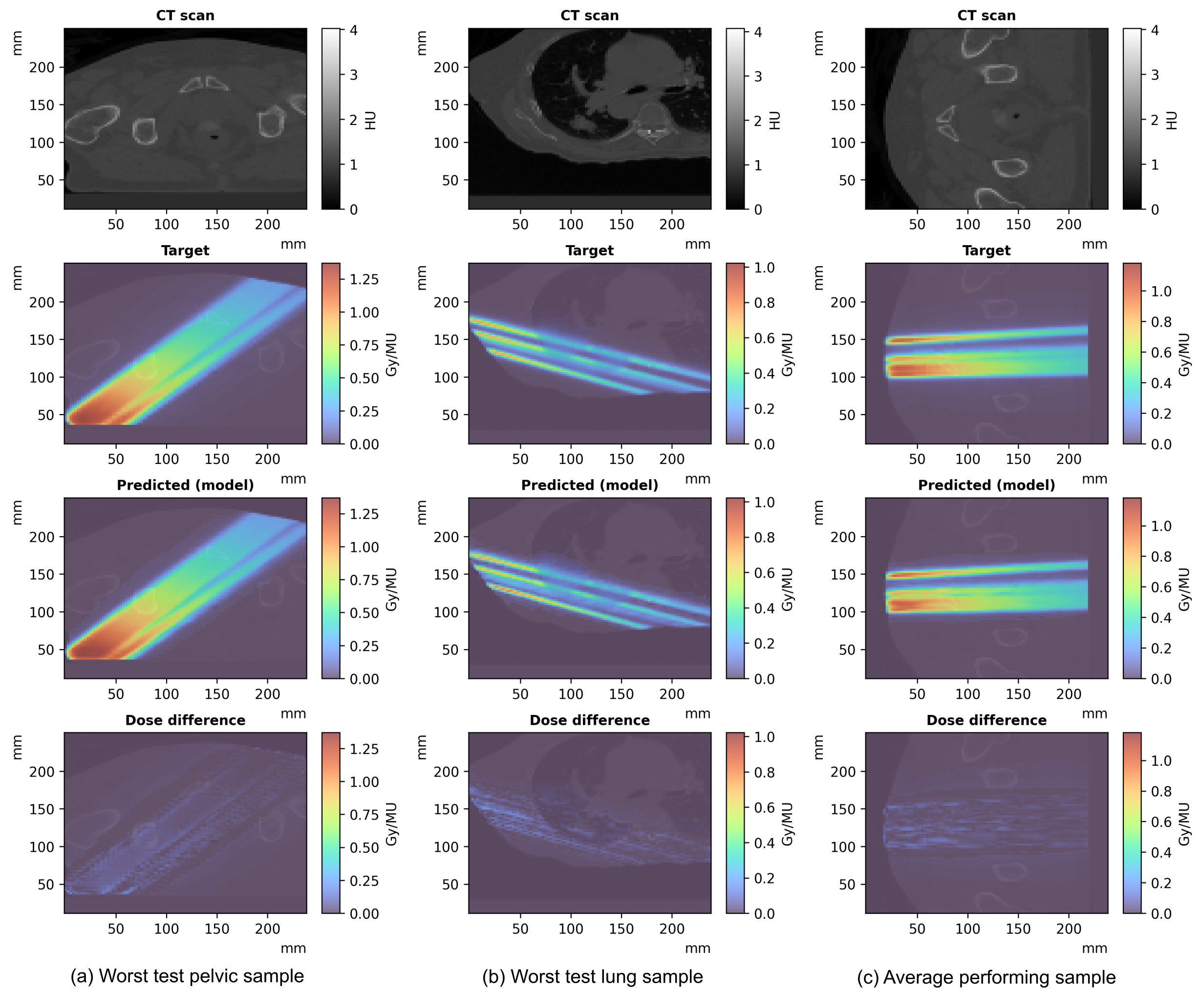}
	\caption{\textbf{Individual beam test samples.} (a) Worst performing pelvic test sample in the gamma evaluation, with $\Gamma(2\text{ mm},2\%)$ gamma pass rate of 87.7\%; (b) worst performing prediction in the gamma evaluation across the lung test samples, with $\Gamma(2\text{ mm},2\%)$ gamma pass rate of 90.1\%, and (c) average performing sample. All subplots show, from top to bottom rows, the central slice of the 3D input CT grid, the reference dose distribution, the model's prediction and the dose difference between the predicted and reference beams.}
	\label{fig:beams}
\end{figure}

\paragraph{Individual beams.} We compute the $\Gamma(1\text{ mm},1\%)$,  $\Gamma(2\text{ mm},2\%)$ and  $\Gamma(3\text{ mm},3\%)$ gamma pass rate and the error $\rho$ for the 584 beams in the test dataset. In \tabref{beams} we compare the mean, standard deviation and minimum values to those reported in previous studies achieving state-of-the-art performance, i.e, the convolutional architectures for photon dose prediction in standard linear accelerator (Linac) \cite{kontaxis_deepdose_2020} and MR-Linac settings \cite{tsekas_deepdose_2021}. In general, iDoTA achieves better pass rates, with higher means and smaller standard deviations. Most importantly, the minimum gamma pass rate across all test samples is >20\% higher than that of the 3D-U-net based architectures. 

iDoTA can better predict photon beams in pelvic anatomies than in lung scans, which is likely caused by the more heterogeneous nature of lung geometries. \figref{beam_dist} further confirms iDoTA's superiority for the pelvic cases over lung, showing $\Gamma(1\text{ mm},1\%)$,  $\Gamma(2\text{ mm},2\%)$, and $\rho$ distributions with lower lung pass rates and higher errors.  \figref{beams} visually compares the target and predicted beam dose distributions for the worst-performing lung and pelvic samples, and an average-performing pelvic beam. The overall errors are low and mostly occur at the beam lateral falloff, which may be caused by the coarse resolution of the input projected shapes.

\begin{table}[t]
	\centering
	\caption{\textbf{Model accuracy for individual broad beams.} Gamma pass rates for photon beams are computed using 3 different criteria in the gamma evaluation. The reported values, which include the mean, standard deviation (std), and minimum across all test samples from pelvic and lung cancer patients, are compared to other state-of-the-art deep learning models as reported in their respective studies.}
	\begin{tabular}{@{}llcccccc@{}}
		\toprule
		\multirow{2}{*}{\textbf{Treatment site}} & \multirow{2}{*}{\textbf{Model}} & \multicolumn{2}{c}{\textbf{$\bm{\Gamma(1,1)}$ [\%]}} & \multicolumn{2}{c}{\textbf{$\bm{\Gamma(2,2)}$ [\%]}} & \multicolumn{2}{c}{\textbf{$\bm{\Gamma(3,3)}$ [\%]}} \\ \cmidrule(l){3-8} 
		&  & \textbf{Mean$\pm$std} & \textbf{Min} & \textbf{Mean$\pm$std} & \textbf{Min} & \textbf{Mean$\pm$std} & \textbf{Min} \\ \midrule
		\multirow{3}{*}{Pelvic} & 3D U-net \cite{kontaxis_deepdose_2020} & \textbf{89.9$\pm$5.1} & 44.5 & 97.8$\pm$3.0 & 55.2 & 99.4$\pm$2.5 & 62.5 \\
		& 3D U-net \cite{tsekas_deepdose_2021} & 87.6$\pm$8.3 & 47.5 & 97.9$\pm$2.6 & 68.2 & 99.5$\pm$1.0 & 77.5 \\
		& iDoTA (ours) & 89.0$\pm$5.4 & \textbf{66.9} & \textbf{98.1$\pm$1.7} & \textbf{87.7} & \textbf{99.6$\pm$0.5} & \textbf{94.7} \\ \midrule
		Lung & iDoTA (ours) & 84.1$\pm$4.7 & 68.9 & 96.9$\pm$2.0 & 90.1 & 99.2$\pm$0.8 & 94.2 \\ \bottomrule
	\end{tabular}
	\label{tab:beams}
\end{table}

\paragraph{Full dose distributions.} For 11 additional patients outside the training dataset with clinical VMAT plans available, we compare the $\Gamma(1\text{ mm},1\%)$,  $\Gamma(2\text{ mm},2\%)$ and  $\Gamma(3\text{ mm},3\%)$ gamma pass rate to the values reported in previous studies. In particular, iDoTA's accuracy and inference times are compared to those of: convolutional U-net architectures predicting each beam in the plan individually \cite{kontaxis_deepdose_2020, tsekas_deepdose_2021}; convolutional models de-noising MC dose distributions \cite{neph_deepmc_2021, bai_deep_2021}; and a concurrent 3D U-net and transformer model for MR-Linac dose prediction \cite{xiao_transdose_2022}.

\begin{figure}[]
	\centering
	\begin{subfigure}[]{\textwidth}
		\centering
		\includegraphics[width=\textwidth]{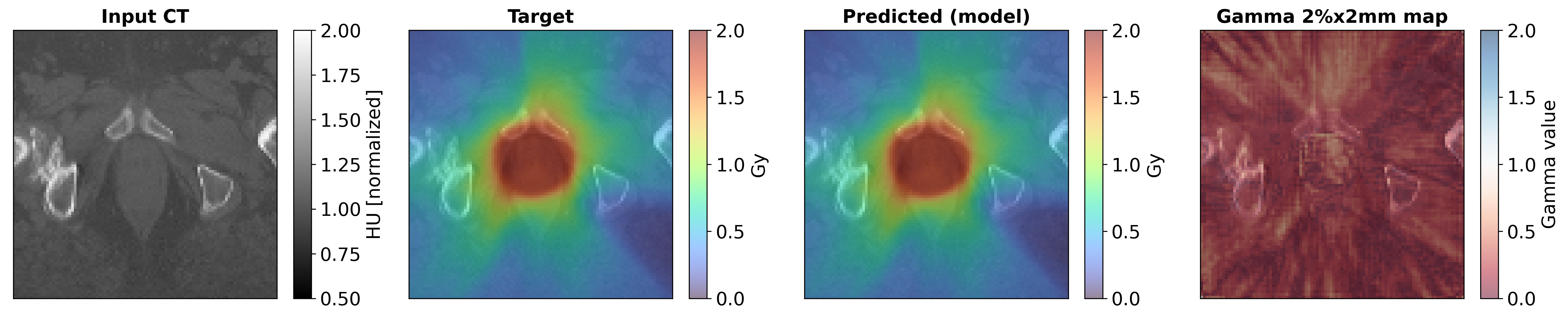}
		\caption{Prostate patient.}
		\label{fig:planP}
	\end{subfigure}
	\begin{subfigure}[]{\textwidth}
		\centering
		\includegraphics[width=\textwidth]{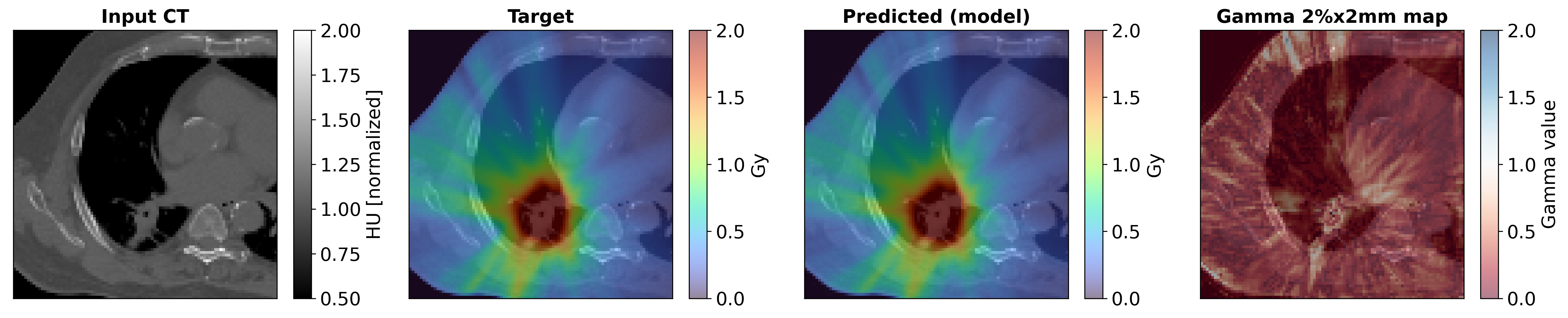}
		\caption{Lung patient.}
		\label{fig:planL}
	\end{subfigure}
	\caption{\textbf{Dose distributions from VMAT plans.} From left to right, the input CT, target and predicted dose distributions and $\Gamma(2\text{ mm},2\%)$ gamma map are shown for two clinical VMAT plans from a (a) prostate and (b) lung cancer patient. To show details of the high dose region, we display crops around the target volume.}
	\label{fig:plans}
\end{figure}

\tabref{plans} shows the mean and standard deviation of the gamma pass rates separately for pelvic, lung and HN patients, comparing them to other models. With a $99.51\pm0.66\%$ (2 mm, 2\%) pass rate, an average relative dose error of $0.75\pm0.36$\% across all patients, and higher pass rates in all treatment sites, iDoTA outperforms all previous approaches. Additionally, the average error $\rho$ in HN, lung and pelvic plans is 1.11\%, 0.64\%, and 0.45\%, respectively. For the remaining patient with a brain tumor, a $\Gamma(1\text{ mm},1\%)$,  $\Gamma(2\text{ mm},2\%)$ and  $\Gamma(3\text{ mm},3\%)$ gamma pass rate of 93.5, 99.7, and 99.9, respectively. As seen in the individual beams, iDoTA is more accurate in pelvic cases and less precise in HN anatomies, although the overall pass rate is still significantly higher than other approaches. Finally, \figref{plans} shows very similar reference and predicted dose distributions for a prostate and lung VMAT plan, along with the corresponding $\Gamma(2\text{ mm},2\%)$ map with mostly all voxels passing the gamma evaluation.

\begin{table}[]
	\centering
	\caption{\textbf{Model accuracy for full clinical dose distributions.} For different treatment sites, we display the gamma pass rates of full photon dose distributions, using 3 different criteria. We include the values from few of the best-performing models as reported in their respective studies. All pass rates include the average and standard deviation across all available dose distributions.}
	\begin{tabular}{@{}llccc@{}}
		\toprule
		\textbf{Treatment   site} & \multicolumn{1}{c}{\textbf{Model}} & \textbf{$\bm{\Gamma(1,1)}$ [\%]} & \textbf{$\bm{\Gamma(2,2)}$ [\%]} & \textbf{$\bm{\Gamma(3,3)}$ [\%]} \\ \midrule
		\multirow{3}{*}{Head \& Neck} & TransDose \cite{xiao_transdose_2022} & - & 96.7$\pm$2.3 & - \\
		& Denoising U-net \cite{neph_deepmc_2021} & 70.9$\pm$2.9 & 89.4$\pm$3.7 & - \\
		& iDoTA (ours) & \textbf{80.5$\pm$8.6} & \textbf{98.9$\pm$0.9} & \textbf{99.9$\pm$0.1} \\ \midrule
		\multirow{5}{*}{Pelvic} & TransDose \cite{xiao_transdose_2022} & - & 97.9$\pm$0.4 & - \\
		& 3D U-net  \cite{kontaxis_deepdose_2020} & 89.9$\pm$3.3 & 99.5$\pm$0.7 & 99.9$\pm$0.3 \\
		& 3D U-net  \cite{tsekas_deepdose_2021} & 82.2$\pm$9.7 & 96.1$\pm$3.1 & 99.4$\pm$0.6 \\
		& Denoising U-net  \cite{bai_deep_2021} & - & 95.4$\pm$1.6 & - \\
		& iDoTA (ours) & \textbf{95.8$\pm$3.1} & \textbf{99.8$\pm$0.2} & \textbf{99.9$\pm$0.0} \\ \midrule
		\multirow{2}{*}{Lung} & TransDose  \cite{xiao_transdose_2022} & - & 96.7$\pm$1.4 & - \\
		& iDoTA (ours) & \textbf{94.3$\pm$1.5} & \textbf{99.8$\pm$0.2} & \textbf{99.8$\pm$0.1} \\ \bottomrule
	\end{tabular}
	\label{tab:plans}
\end{table}

\paragraph{Prediction times.} Computation speed is critically important in adaptive workflows. In \tabref{time}, we compare iDoTA's total time needed to predict individual beams and full plans to the reported values for models in previous studies. All prediction times for all models include the time needed to generate and prepare the inputs, predict the output and (for full dose distributions) accumulate beam doses. For individual beam prediction, iDoTA is significantly faster than any other competitor, being 30-60x faster than the 3D U-net models and 6x faster than the concurrent transformer model TransDose \cite{xiao_transdose_2022}. Likewise, iDoTA predicts full dose distribution from VMAT plans (with 194-354 beams per plan) on average in 8 seconds, representing a 10-80x speed-up compared to the IMRT (with $\approx$ 10 beams) U-net models.  

\begin{table}[]
	\centering
	\caption{\textbf{Average prediction time.} For the fastest models in literature, we report the average computing time needed to predict a photon beam or full dose distribution. The reported values include the time needed to generate and process the model inputs.}
	\begin{tabular}{@{}llc@{}}
		\toprule
		& \textbf{Model} & \textbf{Average time [ms]} \\ \midrule
		\multirow{4}{*}{Photon beams} & TransDose \cite{xiao_transdose_2022} & 310 \\
		& 3D U-net \cite{tsekas_deepdose_2021} & 3000 \\
		& 3D U-net \cite{kontaxis_deepdose_2020} & 1500 \\
		& iDoTA (ours) & \textbf{50} \\ \midrule
		& \textbf{Model} & \textbf{Average time [s]} \\ \midrule
		\multirow{4}{*}{Full plans} & Denoising U-net \cite{bai_deep_2021} & 150 \\
		& Denoising U-net \cite{neph_deepmc_2021} & 660 \\
		& 3D U-net \cite{kontaxis_deepdose_2020} & 60 \\
		& iDoTA (ours) & \textbf{8} \\ \bottomrule
	\end{tabular}
	\label{tab:time}
\end{table}

\section{Discussion}
\label{sec:Discussion}
\paragraph{Comparison to previous models} Framing photon dose calculation as sequence modeling, iDoTA is able to predict beam doses with high accuracy and speed, achieving an overall $97.72\pm1.93\%$ $\Gamma(2\text{ mm},2\%)$ pass rate in lung and pelvic geometries. This per-beam prediction precision translates into a very high $\Gamma(2\text{ mm},2\%)$ pass rate of $99.51\pm0.66\%$ in dose distributions from clinical VMAT plans, which also outperforms all previous models. Compared to the best-performing convolutional models \cite{kontaxis_deepdose_2020, tsekas_deepdose_2021}, iDoTA offers more than 30x faster beam dose prediction even in the most heterogeneous geometries, achieving better gamma pass rates on average with lower spread, and 20\% higher pass rates in the most difficult samples. Furthermore, iDoTA only uses the 3D CT and beam intensity to predict doses, in contrast to the 5 different input volumes containing physics information required by the 3D U-nets, allowing for lower input generation times and faster calculation times overall.
iDoTA also convincingly outperforms MC de-noising models \cite{neph_deepmc_2021, bai_deep_2021}, with a 5-10\% increase in gamma pass rates and a 20-80x speed-up, partially caused by the time needed to generate the high-noise MC dose inputs. Moreover, our method outperforms the concurrent TransDose transformer model in both accuracy and speed. Although TransDose is trained to predict photon beams under magnetic fields for MR-Linac applications --- which could be a more difficult task to learn --- we hypothesize that part of iDoTA's success is due to differences in the model, i.e., that the data-demanding transformer architecture in iDoTA routes information only between each of the 96 slices, instead of the thousands of pixels in TransDose. As a result, iDoTA's transformer has less parameters, which can be favorable with smaller datasets and accelerates inference.

With higher accuracy and lower computing times than any other previously introduced deep learning model, the proposed iDoTA represents a new state of the art in data-driven photon dose calculation. iDoTA can predict full dose distributions in 6-10 seconds, including CT cropping and rotation time ($\approx 25$ ms per beam), ray tracing input calculation ($\approx 0.1$ ms per beam) loading the model and weights ($\approx 2$ s), inferring the beam dose distribution ($\approx 20$ ms per beam) and accumulating the doses in the final grid ($\approx 5$ ms per beam). As a result, iDoTA is an order of magnitude faster than clinically used algorithms or MC approaches adapted to GPU hardware \cite{jia_gpu-based_2011,hissoiny_fast_2011,jahnke_gmc_2012}. While such MC-GPU implementations are several orders of magnitude faster and almost as accurate as their CPU counterparts, their total calculation times are still in the order of minutes. Furthermore, iDoTA is 20x and 60x faster than the Eclipse Acuros XB and AAA algorithms (Varian Medical Systems) used in $\approx 80\%$ of the clinics, which predict VMAT doses in 2-3 and $\approx 10$ minutes, respectively \cite{yan_clinical_2017, fogliata_critical_2012}. Most importantly, the photon beams can be predicted in parallel in several batches depending on the number of GPUs and their internal memory, practically allowing for further reduction in total calculation times.

\paragraph{Limitations} Like all other data-driven algorithms, iDoTA is trained to emulate dose distributions from a specific machine and settings. Deep learning algorithms have limited extrapolation capabilities outside the training domain, which would require a different model each time the machine configuration is changed (or even the CT scanner, unless different CT machines are included in the training dataset). In such cases, fine-tuning iDoTA starting from the provided weights using a smaller dataset can save time without significantly degrading performance.

Ideally, all machine characteristics would be given to the model as separate inputs. Alternatively, to account for geometrical information and machine characteristics, iDoTA requires the additional input projected shape, necessitating ray-tracing pre-calculations. As for the machine parameters, such beam information could be included in the input as separate tokens, e.g., the aperture shape could be given as 2D binary mask at the beginning of the input sequence.

iDoTA is trained using a certain resolution and grid dimensions, which must be fixed for both training and inference. For dose prediction in finer grid resolutions, iDoTA can be coupled to neural representation models capable of accurate super-resolution \cite{vasudevan_implicit_2022}. Regarding grid size, predicting dose distributions from treatment plans or beams through anatomies larger than the predetermined voxel grid must be done in steps, obtaining several input volumes and accumulating the outputs along the beam depth. Conversely, all doses can be predicted for the same fixed grid covering the part of the anatomy containing the structures of interests, which neglects the (usually) low doses near patient entrance. As observed in proton dose prediction \cite{pastor-serrano_learning_2022, pastor-serrano_millisecond_2022}, we expect iDoTA to perform equally well for different grid settings, with calculation times going up for larger grids and finer resolutions, but still within sub-second speed. 

\paragraph{Applicability} Conditioned only on the beam shape projection and the CT, iDoTA is a versatile algorithm that can drastically reduce computing times in any application involving repeated calculation of dose distributions, e.g., checking plan robustness by quickly predicting the dose in each of the many possible error scenarios or anatomical variations of the patient \cite{tilly_dose_2017}. Given a pre-treatment CT, iDoTA can allow fast quality assurance by comparing the estimated and planned dose distributions, with potential applications in online adaptive workflows. Most critically, iDoTA provides the millisecond speed needed in real time adaptive treatments, which can be further reduced if pre-computing all beam shape intensity volumes for each angle in the treatment plan. Future work could even include the magnetic field strength as an additional token in the sequence, similar to the energy token in previous transformer-based proton dose prediction models.

\section{Conclusion} 
Combining the convolutional layers extracting local features with a Transformer backbone routing distant information, iDoTA outperforms any previous deep learning model in photon dose calculation. The presented iDoTA model predicts beam dose distributions in few milliseconds with high accuracy. The per-beam prediction speed translates into estimating full VMAT dose distributions in less than 10 seconds on average, instead of the several minutes required by clinical algorithms or previous data-driven models. Given its speed and versatility, iDoTA can accelerate several steps of the radiotherapy workflow: from treatment planning and quality assurance to real-time adaption.

\section*{Acknowledgments}
This work is supported by KWF Kanker Bestrijding [grant number 11711] and is part of the KWF research project PAREL. Zolt\'an Perk\'o would like to thank the support of the NWO VENI grant ALLEGRO (016.Veni.198.055) during the time of this study. Lei Xing wishes to acknowledge the supports of the National Institutes of Health (NIH) (1R01CA223667, 1R01CA176553, and 1R01CA227713) and Varian Medical Systems (Palo Alto, CA).

\section*{Code availability}
The code, weights and results are publicly available at \url{https://github.com/}.

\section*{CRediT authorship contribution statement}
\textbf{Oscar Pastor-Serrano}: Conceptualization, Methodology, Software, Validation, Formal Analysis, Investigation, Data Curation, Writing – original draft, Visualization. \textbf{Peng Dong}: Conceptualization, Methodology, Formal Analysis, Resources, Data Curation, Writing – Review \& editing. \textbf{Charles Huang}: Conceptualization, Resources, Data Curation, Writing – Review \& editing. \textbf{Lei Xing}: Conceptualization, Methodology, Resources, Writing – Review \& editing, Supervision, Funding Acquisition. \textbf{Zolt\'an Perk\'o}: Conceptualization, Methodology, Formal Analysis, Resources, Writing – Review \& editing, Supervision, Project Administration, Funding Acquisition.

\small
\bibliographystyle{bibstyle}
\bibliography{dota_journal}

\normalsize

\end{document}